\documentclass[journal]{IEEEtran}
\pdfoutput=1
\usepackage{cite}
\ifCLASSINFOpdf
\else
\fi
\usepackage{amstext}

\usepackage{graphicx}
\usepackage{color}

\usepackage[latin1]{inputenc}

\begin{document}
\title{Reconfigurable Waveguides Using Glide-Symmetric Bed of Nails: Design of an All-Metal Switch at Millimetre-Wave Band}

\author{Mohammad~Bagheriasl, Julien~Sarrazin,~\IEEEmembership{Senior Member, IEEE} and Guido Valerio,~\IEEEmembership{Senior Member, IEEE}% <-this % stops a space
\thanks{Mohammad Bagheriasl, Julien Sarrazin and Guido Valerio are with Sorbonne Universit\'{e}, Laboratoire G\'{e}nie \'{e}lectrique et \'{e}lectronique de Paris, F-75005 Paris, France (e-mail: mohammad.bagheriasl@sorbonne-universite.fr, guido.valerio@sorbonne-universite.fr).}% <-this % stops a space
\thanks{This work has been submitted to the IEEE for possible publication. Copyright may be transferred without notice, after which this version may no longer be accessible. This work was supported by the French governement under the ANR grant HOLeYMETA ANR JCJC 2016 ANR-16-CE24-0030 and by Sorbonne Universit\'{e}s under the EMERGENCE 2016 grant MetaSym. This work was performed within the NOVIS60 project supported by the CEFPRA (Indo-French Center for the Promotion of Advanced Research). This publication is based upon work from COST Action Symat (CA18223), supported by COST (European Cooperation in Science and Technology)}}

\maketitle

\begin{abstract}
%Glide-symmetric metasurfaces are used to implement electromagnetic band-gap materials having high stop-band attenuation. They have also found applications in lens antennas thank to their capability to support an almost dispersion-less wave propagation. In this paper, we propose a reconfigurable millimeter-wave artificial waveguide using glide-symmetric pin-like contact-less metasurfaces. The meta-structures are used both as EBG material to confine the field in the guide and as low-dispersive guiding medium filling the guide itself. We enable a two-state reconfigurability by adjusting the displacement between the metasurfaces. A higher displacement enables the wave propagation inside the waveguide and a lower displacement suppresses it. We introduce a proper matching mechanism and an appropriate feeding system. The effect of metal losses in the proposed technology is discussed and the isolation between two adjacent waveguides is computed.
A reconfigurable millimeter-wave artificial waveguide using glide-symmetric pin-like contact-less metasurfaces is proposed in order to enable low loss and high-power-handling capability in switching operations. Thanks to the remarkable behaviour of glide symmetry, the meta-structures are used both as wide-band EBG material to confine the field in the guide and as low-dispersive guiding medium filling the guide itself. The reconfigurability is achieved by adjusting the displacement between the metasurfaces. A higher displacement enables propagation within the waveguide and a lower displacement suppresses it. A two-state reconfigurability is therefore designed, allowing the device to act as an on-off switch in the V-band. In particular, a dispersion analysis highlights the potential of such technology to achieve switching operating over a 55-66~GHz frequency range. A complete design including matching mechanism and WR15 feeding is also described and shown to exhibit a $S_{11}<-10$~dB bandwidth from 57.4 to 62.8~GHz. In the ``off" state, the switch isolation is better than 65~dB and in the ``on" state, the insertion losses are better than 0.7 dB within the entire operating frequency range. The isolation between two adjacent waveguides is also discussed to assess the integration of such a device within a switching network. 
\end{abstract}


% \fi
\IEEEpeerreviewmaketitle

\section{Introduction}
\label{sec:introduction}

\IEEEPARstart{F}{ifth-generation} (5G) of mobile communication systems addresses the ever-growing need of higher throughput wireless connectivity in our society \cite{andrews2014will,boccardi2014five,hong2017multibeam}. High data rate of the order of gigabits per second (Gb/s), a latency time in milliseconds, high traffic volume density, and improved spectral and energy efficiencies are the requirements for this new generation of wireless communications. Millimeter-wave (MMW) wireless systems are emerging as a promising technology for achieving these requirements \cite{pi2011introduction,qiao2015enabling,roh2014millimeter,rappaport2013millimeter}. This is due to the vast amount of underutilized spectrum in MMW frequencies along with the recent advances in designing low-cost high-performance integrated circuits (ICs) at such frequencies \cite{razavi2009design,doan2005millimeter,chen2012silicon}. So, frequency bands such as 28~GHz and 40~GHz have been ratified for 5G use while the 3GPP release 17 is currently investigating the spectrum above 52.5~GHz, with a specific focus on the 60~GHz licence-free band \cite{3gpp}. MMW multibeam antennas (MBAs) are considered in the phase 2 of 5G deployment with suitable beamforming techniques to achieve massive multiple-input, multiple-output (MIMO) \cite{swindlehurst2014millimeter,larsson2014massive,bogale2016massive}, thus significantly enhancing the spectral and energy efficiency while minimizing interference levels.

Radio frequency (RF) chains, digital-to-analog converters (DACs), and analog to digital converters (ADCs) are the most expensive parts of a wireless transceiver. In MMW massive MIMO wireless systems, due to the huge number of antennas to mitigate large free-space attenuation, employing one RF chain and one ADC or DAC for each antenna is not feasible as doing such also increases the energy consumption of the transceiver to an unacceptable level. Hybrid analog/digital beamforming architectures are proposed as a solution to this limitation \cite{alkhateeb2013hybrid,alkhateeb2014channel,bogale2014beamforming,han2015large}. % In these structures, the MIMO processing is executed partly in digital and in analog domains to reduce the number of RF chains.
In such architectures, the number of RF chains is typically much lower than the number of antennas and therefore, analog beamformers with phase shifters \cite{stanley2017high}, Butler Matrices \cite{wang2019compact}, or lenses~\cite{brady2013beamspace}, are considered to feed all antennas. While Butler-Matrix-based and lens-based hybrid arrays exhibit less complexity, they require a beam selector made of switches to map the few RF chains to the large number of analog beamformer inputs
~\cite{brady2013beamspace}. %where a beam selector is used to switch between different predefined antenna beams , implemented with phase shifters.
Recently, switching networks have also been suggested to replace phase shifters in analog beamformers in order to reduce complexity and increase power and spectral efficiencies \cite{mendez2016hybrid}. Consequently, in this context, the design of power-efficient millimeter-wave switches is a key aspect in the implementation of hybrid analog/digital massive MIMO systems.

Integrated circuits at microwave and millimeter-wave frequencies can be conveniently realized with gap-waveguide technology \cite{kildal2008local}. In this technology, the electromagnetic wave is guided in an air region rather than in a dielectric, resulting in reduced losses compared e.g. to substrate-integrated waveguides. Moreover, the metallic top and bottom plates of gap waveguides are contactless. This facilitates the manufacturing process which can be performed in two separate parts. Gap waveguides are designed by mimicking a perfect magnetic conductor (PMC) boundary condition on the two sides of the waveguide by means of an electromagnetic band-gap (EBG) material \cite{rajo2011numerical,kildal2011design, vosoogh2015corporate}. Bed-of-nail metasurface is among the most widely used configurations for this purpose \cite{silveirinha2008electromagnetic,maci2011metasurfing}. 

Recently, new EBG designs for gap waveguides have been proposed in order to enhance their confinement properties without adding significant complexity to the design \cite{ebrahimpouri2017design,8093742}. These solutions are based on the use of two metasurfaces on both plates of the waveguide, having a mutual shift of half  period in both periodicity directions. This symmetry configuration is commonly called glide symmetry. It is a higher-order symmetry of periodic structures \cite{glide,Crepeau_1964,Mittra,Kieburtz_1970} which has recently gained interest thanks to discovering new interesting dispersive behaviour \cite{bagheriasl2019bloch}. On one hand, they are less frequency-dispersive than their non-glide-symmetric counterparts in their first pass-band, leading to their adoption in design of ultrawideband lenses \cite{quevedo2015ultrawideband, Lens_Ericsson_2,7928545}. On the other hand, they have wider and stronger stop-bands compared to non-glide-symmetric periodic structures \cite{Mahsa_flanges, Kexin_15}, explaining why they are used in EBGs for gap waveguides.
These two properties, together with the contactless nature of the metasurfaces, make glide-symmetric metasurfaces particularly interesting to embed reconfigurable properties in gap waveguides. The impact of some geometric parameters on the transmission properties of glide-symmetric waveguides was studied in recent work \cite{Pin_symmetry_Eva}. A phase-shifter has been proposed in  glide-symmetric gap waveguide technology where the phase shift is adjusted by changing the height of a dielectric slab that is inserted inside the waveguide \cite{Eva_Phaseshifter}. %its structure as a means to reconfigure the phaseshifter. Interchanging discrete number of dielectric slabs while reassembling the structure has been proposed as another way to reconfigure the values of phase shift.
In \cite{Padilla_2018} different kinds of pin-like structures inside a waveguide lead to different phase shifts.

In this paper, we propose an integrated waveguide presenting a two-state functioning. An ``on-state'' allows the propagation through the guide, while an ``off-state'' stops the propagation. This reconfiguration is done by mechanically adjusting the distance between the top and bottom plates, which is made possible thanks to their contact-less configuration. Glide-symmetric EBGs offer strong stop-band characteristics, and glide-symmetric pins inside the waveguide ensure nearly dispersion-less wave propagation.

%structure that includes glide-symmetric pins on parallel plates with no physical contact similar to gap waveguide technology. More importantly, we will show that the inclusion of these periodic pins in the guiding medium enables a two-state "on" or "off" waveguide design. The waveguide guides the wave in the "on" state and attenuates the wave in the "off" state.  This can be useful in design of the feeding network for multibeam antennas since it can reconfigure the propagation path of a signal in the switch that feeds different input ports of the antenna \cite{quevedo2015ultrawideband}.

The paper is organized as follows. In Sec.~\ref{sec:waveguide1}, we first present the design concept of the two-state waveguide by choosing the proper EBG and guiding pin configurations. We then introduce the feeding and the impedance matching networks in the proposed waveguide. In Sec.~\ref{sec:Simulation_Results}, we discuss the simulation results of the designed waveguide. First, we investigate the results in a single waveguide. Then, we consider two coupled lines and validate the "on" and "off" states in these waveguides. We also show the good isolation between the two lines using the simulation results.

\section{Reconfigurable Artificial Waveguide}
\label{sec:waveguide1}
In this section, we propose a structure to implement a reconfigurable artificial waveguide that can alternate between ``on'' and ``off'' states. It lets the wave propagate in its ``on'' state, whereas it attenuates the wave while in the ``off'' state. 

\begin{figure}[!tbp]
    \includegraphics[width=\columnwidth]{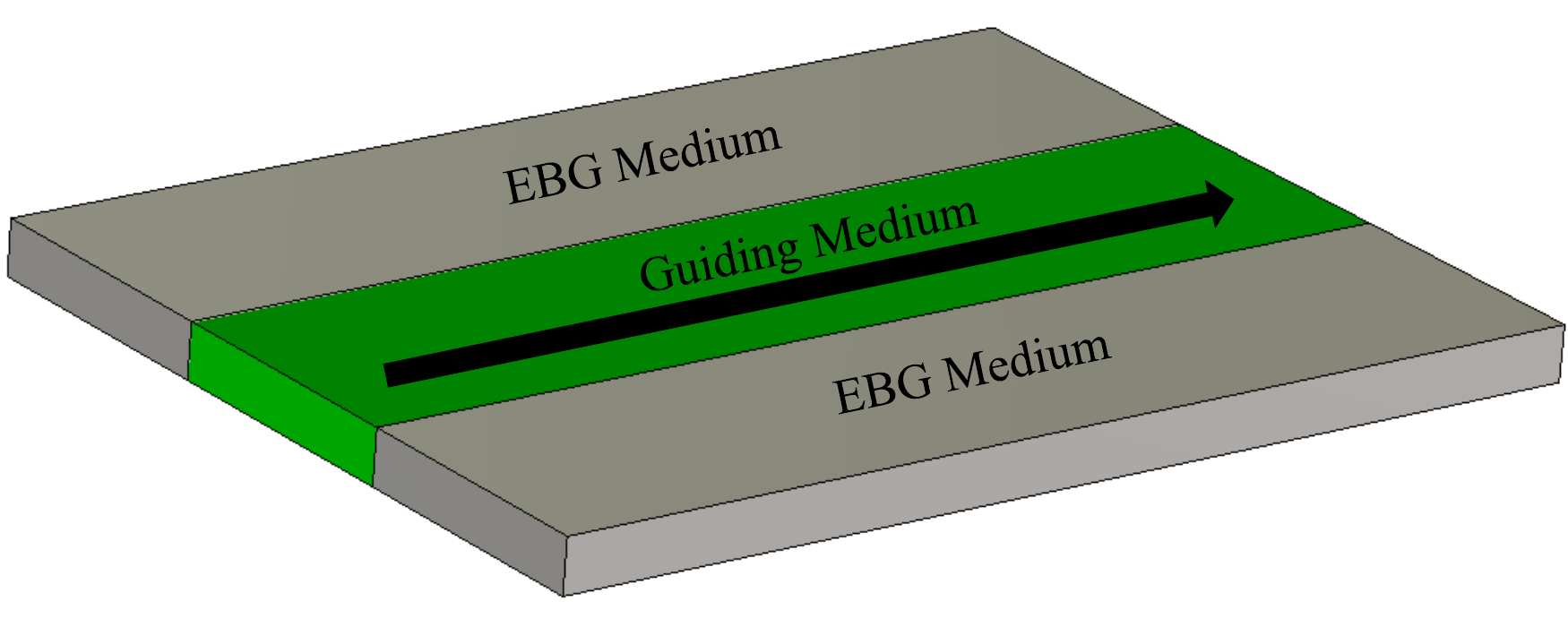}
	\caption{Division of medium regions in an artificial waveguide.}
	\label{fig:Fig0}
\end{figure}

The artificial waveguide employs EBG structures at its lateral boundaries to confine the field inside the guide, which is filled with an artificial medium. Fig.~\ref{fig:Fig0} shows a perspective view of the artificial waveguide with its different medium regions, where the propagation direction is shown with an arrow. The EBG medium confines the field within the guiding medium. We discuss the design of different parts of the waveguide, including the guiding medium, the EBG, and a matching section.
%We will show that the EBG structures enforce a lateral confinement similar to a PMC boundary condition. \textcolor{yellow}{We did not show it. Should I delete modify this phrase and drop the similarity to PMC boundary condition?} 

\subsection{Design of Guiding Medium and EBG}
\label{subsec:EBG_guiding}
\begin{figure}[!tbp]
    \includegraphics[width=\columnwidth]{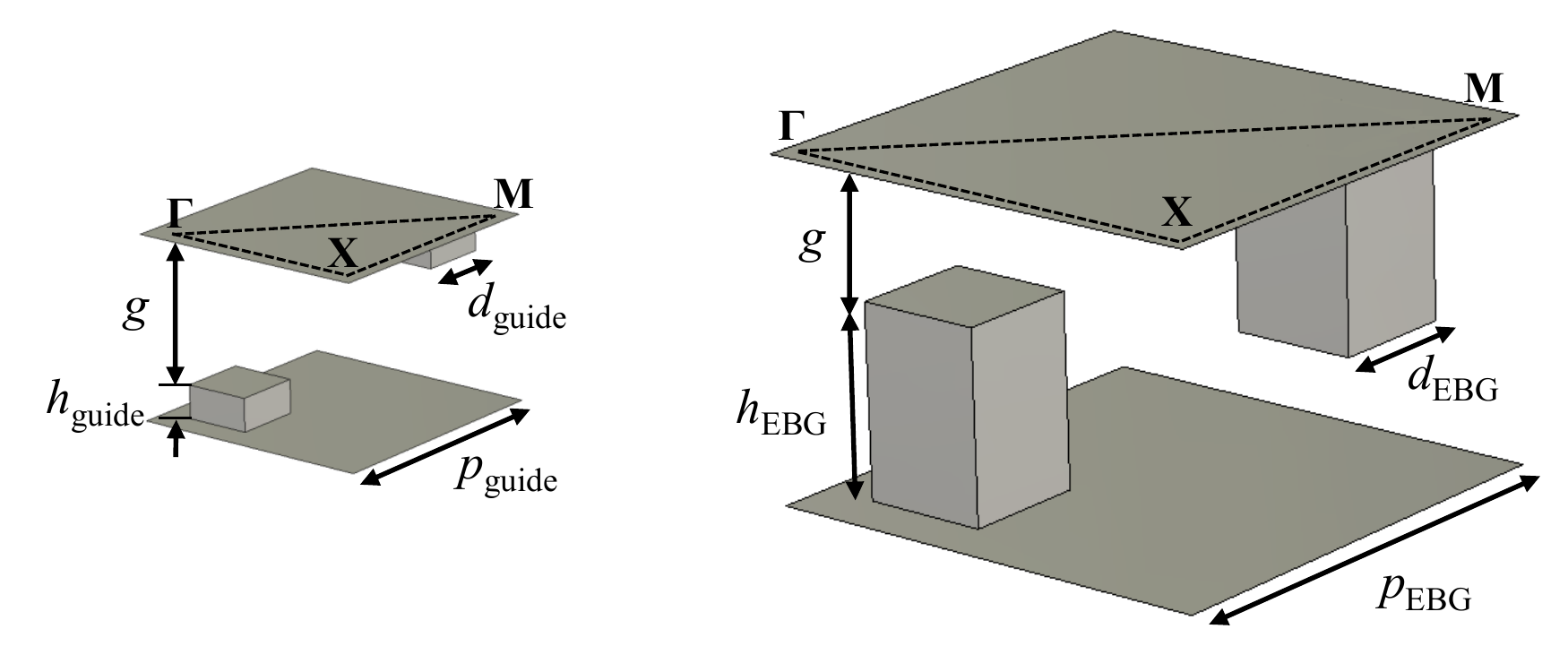}
	\caption{Unit cells of the glide-symmetric pin-like mediums used in the artificial waveguide. The unit cell on the left hand side is for the guiding medium and the one on the right hand side is for the EBG. The unit cells are shown with the correct ratio to highlight the differences in their geometrical parameters.}
	\label{fig:Fig1}
\end{figure}

 In order to implement the EBG regions, we use two parallel plates loaded with pins as our waveguide structure. The EBG performance (namely, a high attenuation in the operating frequency range of interest) should ensure that the wave is well confined in the guiding region. We use a glide symmetric configuration of pins in the EBGs to achieve a stronger and wider stopband compared to a single metasurface as suggested in \cite{ebrahimpouri2017design}. The guiding medium of the waveguide is also realized with a pin-like pair of metasurfaces whose geometric parameters are different from those used for the EBG. The unit cells of both EBG and guiding media are shown in Fig.~\ref{fig:Fig1}. Different geometrical parameters lead to different electromagnetic characteristics in the two regions. It is important to note that the only parameter which is kept the same for EBG and guiding medium is the gap $g$ between the tip of a pin and the opposite plate.

%The pin-like plates define both the EBG and the guiding medium according to the geometric parameters.
 
 %In addition to EBGs, we apply glide symmetry to the guiding medium. we will later demonstrate how the inclusion of these periodic pins in the guiding medium will permit the transition between the "on" and "off" states. We choose a glide-symmetric bed of nails to design the EBGs and the guiding medium.

\begin{figure}[!tbp]
    \includegraphics[width=\columnwidth]{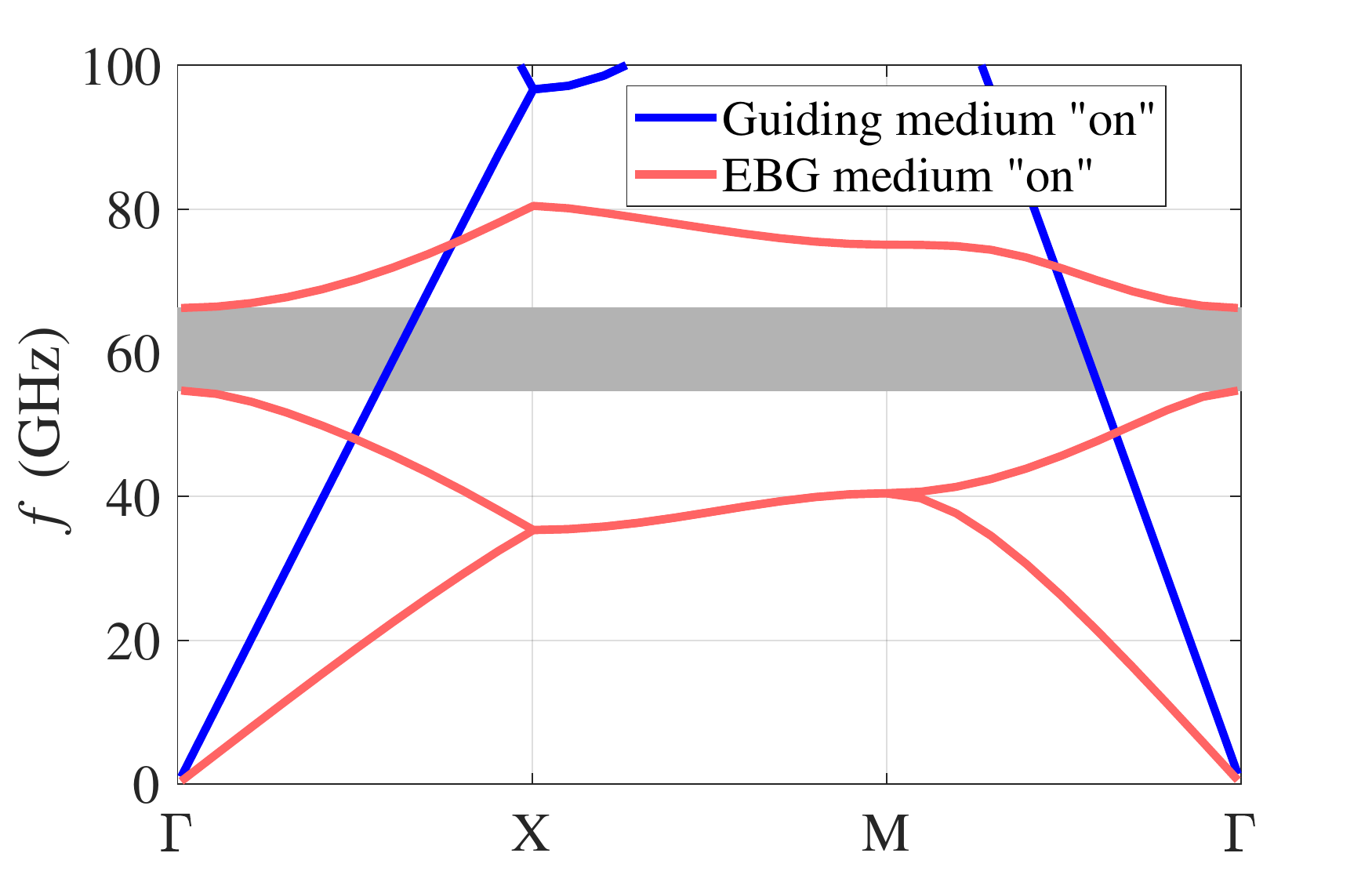}
	\caption{Full dispersion diagram of the structure with the unit cells shown in Fig.~\ref{fig:Fig1} in the ``on" state. Geometrical parameters of the guiding medium: $p_{\mathrm{guide}}=1.5$~mm, $d_{\mathrm{guide}}=0.4$~mm, $h_{\mathrm{guide}}=0.2$~mm and $g=0.9$~mm. Geometrical parameters of the EBG medium: $p_{\mathrm{EBG}}=3$~mm, $d_{\mathrm{EBG}}=0.8$~mm, $h_{\mathrm{EBG}}=1.2$~mm and $g=0.9$~mm.}
	\label{fig:Fig2}
\end{figure}

For an ``on'' state in the waveguide, we design the periodic structure in the guiding region to be in its pass-band, hence able to propagate an electromagnetic wave. We also make certain that the periodic structure in the EBG medium is in its stop-band, thus attenuating the fields outside the guiding medium. %\textcolor{blue}{As it was suggested earlier,} we accomplish this behaviour by using different parameters in the EBG and the guiding region.
%In this paper, we design our waveguide to operate in the band \textcolor{blue}{57-63} GHz (V frequency band). WELL: it is what has been obtained, ideally we would like 50-75GHz (entire V-band). So I would not say it like that.
Fig.~\ref{fig:Fig2} plots the full dispersion diagrams of the unit cells shown in Fig.~\ref{fig:Fig1}. The eigenmode solver of CST Microwave Studio is used to find the dispersion diagrams \cite{CST}. The geometrical parameters of the unit cell of guiding medium are $p_{\rm{guide}}=1.5$~mm, $h_{\rm{guide}}=0.2$~mm, $d_{\rm{guide}}=0.4$~mm and $g=0.9$~mm. The parameters for EBG unit cell are $p_{\rm{EBG}}=3$~mm, $h_{\rm{EBG}}=1.2$~mm, $d_{\rm{EBG}}=0.8$~mm, and $g=0.9$~mm. 
%A comparison of the two diagrams clarifies which of the two unit cells can be used in the guiding medium and which one can confine the wave as an EBG material.
The potential operating frequency range is highlighted by a gray rectangle in Fig.~\ref{fig:Fig2} and covers the 55-66~GHz bandwidth. % Indeed, within this range, one of the two curves displayed indicates that its corresponding medium exhibits a pass-band behavior while the other curve shows that its corresponding medium achieves a stop-band in all possible directions of propagation. This means that we can use the unit cell corresponding to the first curve in our guiding medium while we use the latter in the EBG medium.}
Indeed, within this range, the guiding medium exhibits a linear-frequency-dependent propagation constant and the EBG medium exhibits a bandgap. The field can therefore propagate while being confined with no dispersion.

\begin{figure}[!tbp]
    \includegraphics[width=\columnwidth]{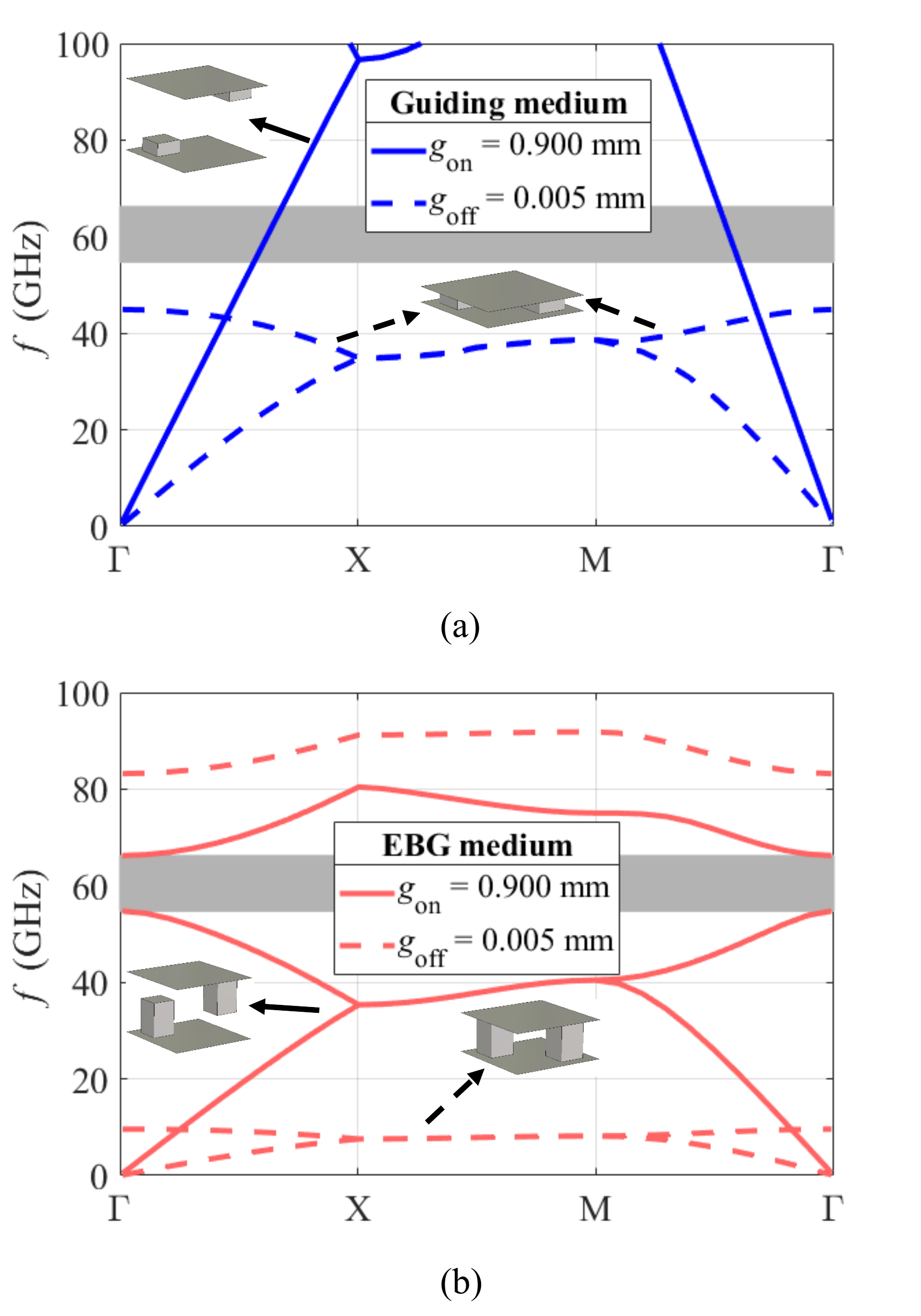}
	\caption{Full dispersion diagram of the unit cell shown in Fig.~\ref{fig:Fig1} (a) for two values of $g=0.005$~mm (off state) and $g=0.9$~mm (on state) in  the guiding medium with geometrical parameters $d_{\rm{guide}}=0.4$ mm, $h_{\rm{guide}}=0.2$ mm and $p_{\rm{guide}}=1.5$ mm (b) for two values of $g=0.005$~mm (off state) and $g=0.9$~mm (on state) in the EBG medium with geometrical parameters $d_{\rm{EBG}}=0.8$ mm, $h_{\rm{EBG}}=1.2$ mm and $p_{\rm{EBG}}=3$ mm.}
	\label{fig:Fig3}
\end{figure}

\subsection{Two-State Reconfigurability}
\label{subsec:reconfigurability}
Thus far, we have properly chosen a guiding medium that can let wave propagate and an EBG medium that can effectively prevent any wave leakage from the guiding region. Now, we realize a transition from the ``on'' (or propagating) state to an ``off'' (or attenuating) state. In the ``off'' state, the EBG medium should still attenuate the fields, thus avoiding wave propagation in the cross-sectional directions like previously. Moreover, the guiding region should become an EBG material itself, within the waveguide operating frequency range, to prevent any wave propagation in the longitudinal direction of the waveguide.

The possibility to switch from the ``on'' to the ``off'' state is offered by the contactless configuration, which allows for adjusting the distance between the two plates. %This may provide a degree of freedom in reconfigurability of the components in this technology as this adjustment can be mechanically applied with little effort.
We investigate the effect of the same variation in the parameter $g$ of the unit cells shown in Fig.~\ref{fig:Fig1}, i.e. the distance between the top of a pin and the opposite metallic plate. This can be easily controlled by moving the two metallic plates either closer to or further away from each other.

Fig.~\ref{fig:Fig3}~(a) depicts the full dispersion diagrams of the guiding medium when varying the gap $g$ of $\Delta g=0.895$~mm. %with parameters of $h=1.2$~mm, $d=0.5$~mm and $p=1.5$~mm 
The two different gaps are $g_{\mathrm{off}}=0.005$~mm and $g_{\mathrm{on}}=0.9$~mm. We notice that by decreasing the gap, %. This means that some frequencies that belonged to the first passband of the guiding medium with $g=0.25$~mm lie in the first stopband of the same medium with $g=0.1$~mm. 
the frequency region highlighted with a gray rectangle (55-66~GHz) passes from an ``on'' state (pass-band) to an ``off'' state (stop-band). More precisely, we can remark that decreasing the gap $g$, the previous pass-band is transformed into a stop-band between the second and third Bloch modes. Any value of $g$ less than 0.007~mm is sufficient to ensure a bandgap in the guiding medium in the ``off" state. %A constraint on the minimum acceptable $g$ is however given by the mechanical precision of the configuration technique. \textcolor{blue}{Another constraint is the frequency coverage of the stopband in the EBG and its variation with respect to changes in the gap $g$.}\textcolor{red}{and this is actually the only constraint to have in mind. So I would remove this sentence too}

Next, we investigate the same gap variation $\Delta g=0.895$~mm on the EBG (varying the gap from $g_{\mathrm{on}}=0.9$~mm to $g_{\mathrm{off}}=0.005$~mm). %We expect that by decreasing the gap, the first stop-band of the EBG medium will widen similar to what happened to the guiding medium. 
Fig.~\ref{fig:Fig3}~(b) displays the full dispersion diagrams of the EBG medium for the two aforementioned gap sizes. %$g=0.1$~mm and $g=0.25$~mm. The other parameters are $h=1.2$~mm, $d=0.5$~mm and $p=2.2$~mm as before. Comparison of the two curves in the figure proves that 
As expected, the smaller the gap gets, the wider the stop-band becomes and the material still behaves as an EBG in the gray-shaded frequency range: the stop-band width is increased for a smaller gap and %the gray region that shows the bandwidth of reconfigurability for the guiding medium (previously shown in Fig.~\ref{fig:Fig3}~(a)) is also sketched here. We observe that 
the EBG medium provides a stop-band in the gray region with both gaps $g_{\mathrm{off}}=0.005$~mm and $g_{\mathrm{on}}=0.9$~mm. Therefore, the design of the ``off'' state is complete. %One important thing to note is that the stop-band of the EBG in its ``on'' state (the larger $g$) is a factor in deciding the minimum $g$ for reconfiguring the waveguide. For instance, too much of a decrease in $g$ from the ``on'' to the ``off'' state may force the gray region highlighted in Fig.~\ref{fig:Fig3}~(a)~and~(b) to exceed the stop-band region of the EBG in its ``on'' state. In this case, part of this region falls in the pass-band of the EBG medium which will fail to confine the wave cross-sectionally at these frequencies. If a wider frequency range of reconfigurability is needed, a better EBG needs to be designed that exhibits a wider stop-band. \textcolor{red}{I think this remark is important since it explains the risk of a wrong choice of the $g$ variation. But I do not understand clearly what you mean. Are talking about the upper part of the gray area which could touch the red curve in Fig. 4(b)? In this case, it is better to explain it this way in just one sentence.} \textcolor{blue}{No, I mean that the bandwidth is firstly limited by the bandwidth of the EBG. So, it is useless to decrease the gap $g$ further if we have a $g_{off}$ that already covers the whole EBG bandwidth. Maybe, we should rephrase it in a more simple way or delete it since it is obvious.}

\subsection{Feeding Mechanism and Impedance Matching}
\label{subsec:feed_matching}
\begin{figure}[!tbp]
    \includegraphics[width=\columnwidth]{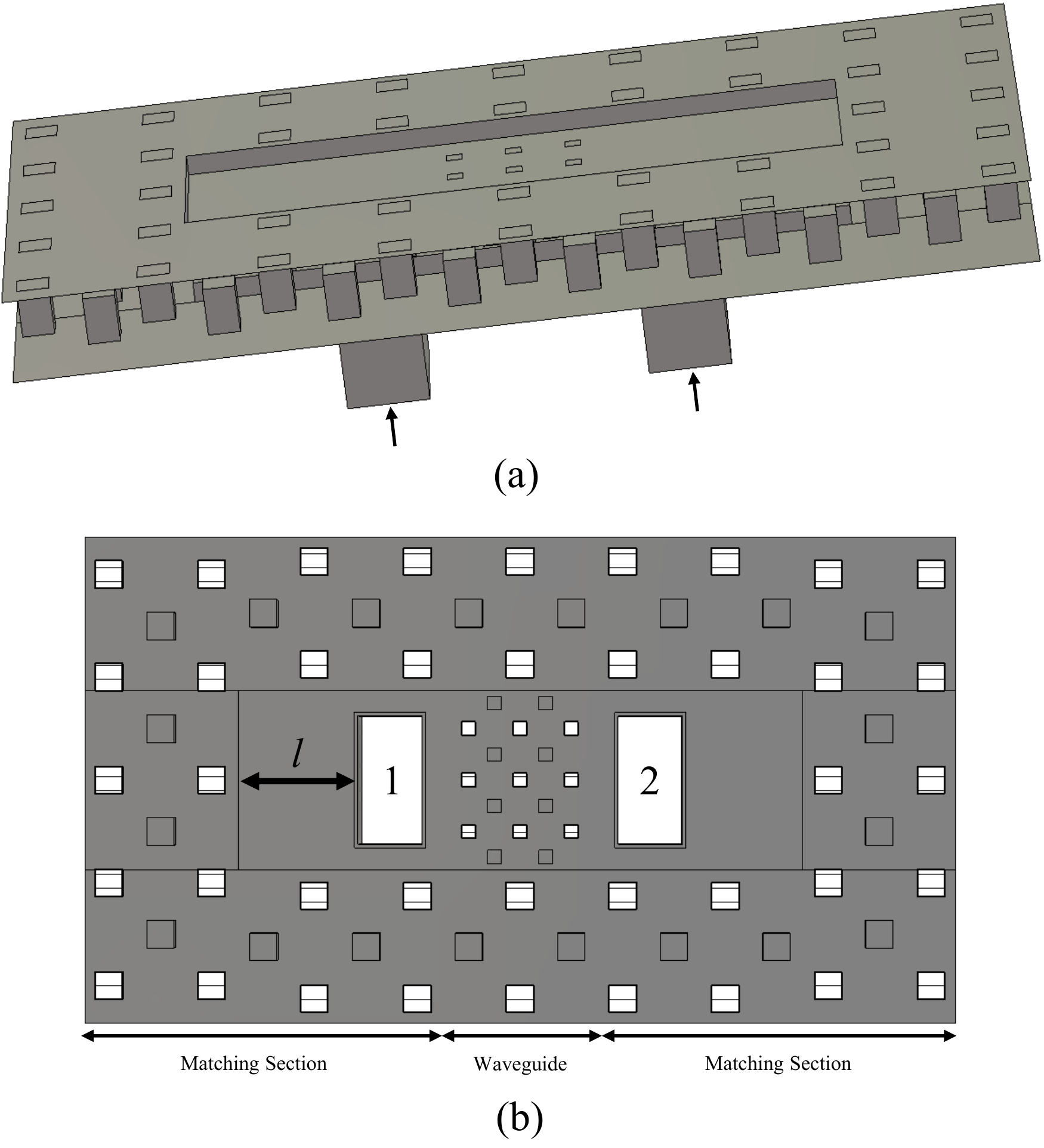}
	\caption{The designed artificial waveguide: (a) perspective view (b) the top view with the top metal plate removed for better visualisation of the inner parts.}
	\label{fig:Fig5}
\end{figure}

%We have already qualified the proper periodic structures for designing the building blocks of an artificial waveguide.
Fig.~\ref{fig:Fig5}~(a) shows the complete proposed waveguide, including the feeding accesses. Two slots of dimensions of 3.7592~mm $\times$ 1.8796~mm are etched in the bottom plate to enable exciting the structure by WR15 standard accesses (V band), normally placed below the waveguide and indicated by the black arrows in the figure. Fig.~\ref{fig:Fig5}~(b) displays a top view of the waveguide with the upper plate removed for a clear visualisation of the pins. Gray squares are pins on the bottom plate, and white squares are the pins locations on the upper plate, removed from the picture. %The waveguide is surrounded by EBG materials in both $x$ and $y$ directions.
The waveguide is divided into three sections: the guiding section in the middle and two matching sections at both ends of the longitudinal direction. Each matching section includes the feeding slot at a distance $l$ from the EBG at one end of the structure. With a proper choice of $l$, the EBG medium reflects the wave from the feed towards the waveguide.

%\begin{figure*}[!tbp]
%    \includegraphics[width=\textwidth]{Fig05.pdf}
%	\caption{The designed artificial waveguide: (a) perspective view (b) the top view with the top metal plate removed for better visualisation of the inner parts.}
%	\label{fig:Fig5}
%\end{figure*}

\section{Simulation Results}
\label{sec:Simulation_Results}
\subsection{Single Waveguide}
\label{subsec:results1}
In this subsection, we discuss the simulation results of the waveguide designed in the previous section. The simulations are performed with the frequency solver of CST Microwave Studio. %\textcolor{blue}{We repeat that the EBG and the guiding mediums are periodic structures with their unit cells shown in Fig.~\ref{fig:Fig1}. The parameters are $p_{\rm{guide}}=1.5$~mm, $h_{\rm{guide}}=0.2$~mm, $d_{\rm{guide}}=0.4$~mm and $g_{\rm{guide}}=0.9$~mm for guiding medium and $p_{\rm{EBG}}=3$~mm, $h_{\rm{EBG}}=1.2$~mm, $d_{\rm{EBG}}=0.8$~mm and $g_{\rm{EBG}}=0.9$~mm for EBG.
We have chosen 7 rows of pins in the guiding medium. This corresponds to a width of $3.5 \times p_{\rm{guide}}$ which equals 5.25~mm and is slightly wider than the length of the rectangular waveguide feed. We have used a width of $1.5{\times}p_{\rm{EBG}}=4.5$~mm for the EBG regions surrounding the waveguide. The length of the structure is 25.5~mm. Furthermore, we empirically found that the distance $l$ providing the best matching bandwidth is 3.4~mm. The metal is modeled as copper. %Thus, the minimum number of pin rows that qualifies is 7. Therefore, we set the width of the guiding medium at the end of the waveguide to $3.5{\times p_1}$ or equally 5.25~mm. This means that a width tapering is required to eliminate the width difference of the guiding medium in different sections of the structure. A 10-step linear width tapering is applied to the structure. Therefore, the width difference from one step to the next is always the same. This part of the matching section is therefore $10{\times}p_2=30$~mm long. We also apply a 20-step linear pin height tapering to the guiding medium. The length of this tapering is $20{\times}p_1=30$~mm and therefore it does not add to the length of the matching. This is due to the fact that both the pin height tapering and the guide width tapering will be applied over the same length as it was already shown in Fig.~\ref{fig:Fig6}~(b).
%Looking at the EBG medium surrounding the waveguide, we have used a width of $1.5{\times}p_1=4.5$~mm for these regions. This corresponds to 3 rows of pins in the EBG medium. Furthermore, we found that the distance $l$ which provides the best matching bandwidth is 2.5~mm.

\begin{figure}[!tbp]
    \includegraphics[width=\columnwidth]{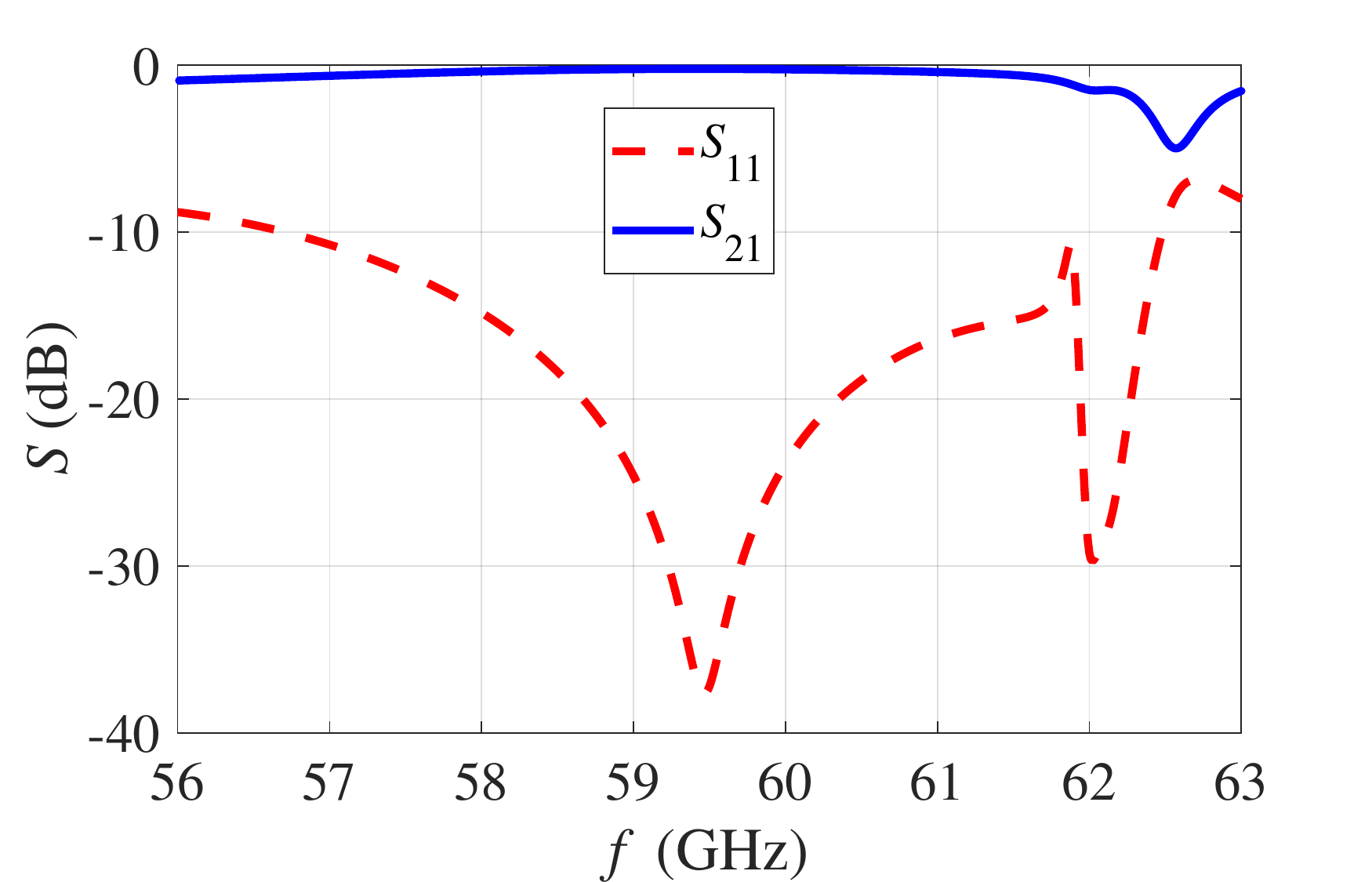}
	\caption{Scattering parameters of the structure shown in Fig.~\ref{fig:Fig5} in the ``on'' state. Geometrical parameters: $l=3.4$~mm, $p_{\rm{guide}}=1.5$~mm, $h_{\rm{guide}}=0.2$~mm $d_{\rm{guide}}=0.4$~mm, $p_{\rm{EBG}}=3$~mm, $h_{\rm{EBG}}=1.2$~mm $d_{\rm{EBG}}=0.8$~mm, $g=0.9$~mm. }
	\label{fig:Fig7}
\end{figure}

\begin{figure}[!tbp]
    \includegraphics[width=\columnwidth]{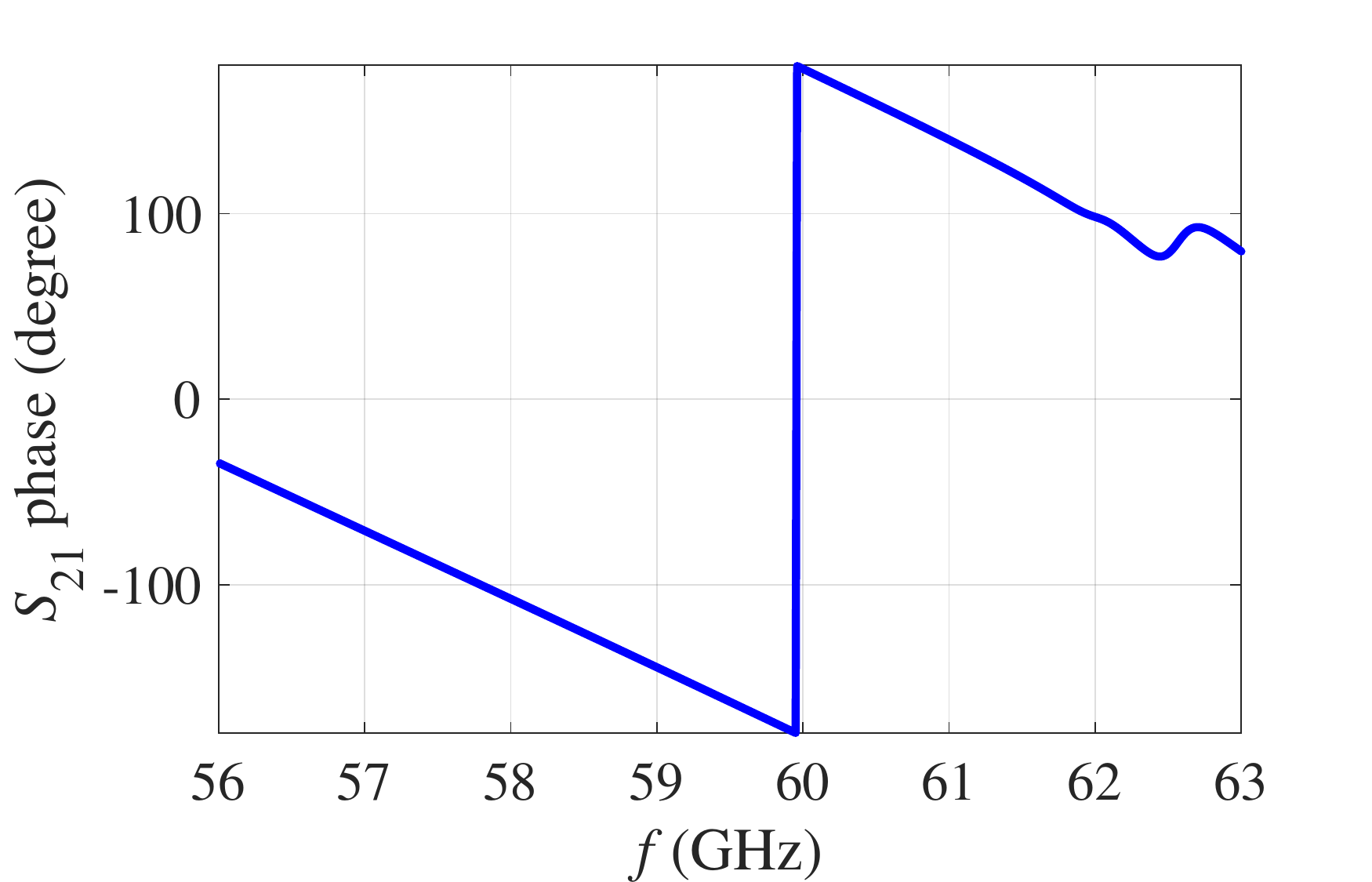}
	\caption{Phase of $S_{21}$ versus frequency for the structure shown in Fig.~\ref{fig:Fig5} in the "on" state. Geometrical parameters: $l=3.4$~mm, $p_{\rm{guide}}=1.5$~mm, $h_{\rm{guide}}=0.2$~mm $d_{\rm{guide}}=0.4$~mm, $p_{\rm{EBG}}=3$~mm, $h_{\rm{EBG}}=1.2$~mm $d_{\rm{EBG}}=0.8$~mm, $g=0.9$~mm. }
	\label{fig:Fig8}
\end{figure}

Fig.~\ref{fig:Fig7} plots the simulated scattering parameters for the designed waveguide with the aforementioned geometrical parameters and the gap $g=g_\text{on}=0.9$~mm (the ``on'' state discussed in the previous section). %Two simulations have been carried out. In the first one the metallic surfaces are made of perfect electric conductor (PEC); in the second one with lossy metal (copper). Both results are shown in the figure. The simulated results with PEC are shown in solid lines.
The $S_{11}$ curve displays a $-10$~dB impedance matching bandwidth of 5.7 GHz between the frequencies $f_{min}=56.8$~GHz and $f_{max}=62.5$~GHz. %\textcolor{red}{you should mention $f_{min}$, $f_{max}$ so that the reader can assess the change in bandwidth due to matching transition}.
The $S_{21}$ curve confirms the full transmission from the first waveguide port at one end to the waveguide port at the other end of the guide. $S_{21}$ levels are above -4~dB in the entire $-10$~dB impedance matching bandwidth. If $S_{21}$ levels of higher than $-0.7$~dB are required, the bandwidth reduces to 4.9~GHz between $f_{min}=56.8$~GHz and $f_{max}=61.7$~GHz. Fig.~\ref{fig:Fig8} plots the phase of $S_{21}$ versus frequency. It is observed that the phase is linear in the aforementioned bandwidth, which is typical of waveguides realized with glide-symmetric geometries. %The simulated results using copper are shown in dashed lines, to study the effect of metal losses. We observe from the $S_{11}$ that the lossy structure still provides the same -10 dB bandwidth as the structure with PEC. The $S_{21}$ curve displays a drop of less than 0.014 dB/{$\lambda_0$},  where $\lambda_0$ is the wavelength of the wave in free space at 60 GHz. 

\begin{figure}[!tbp]
    \includegraphics[width=\columnwidth]{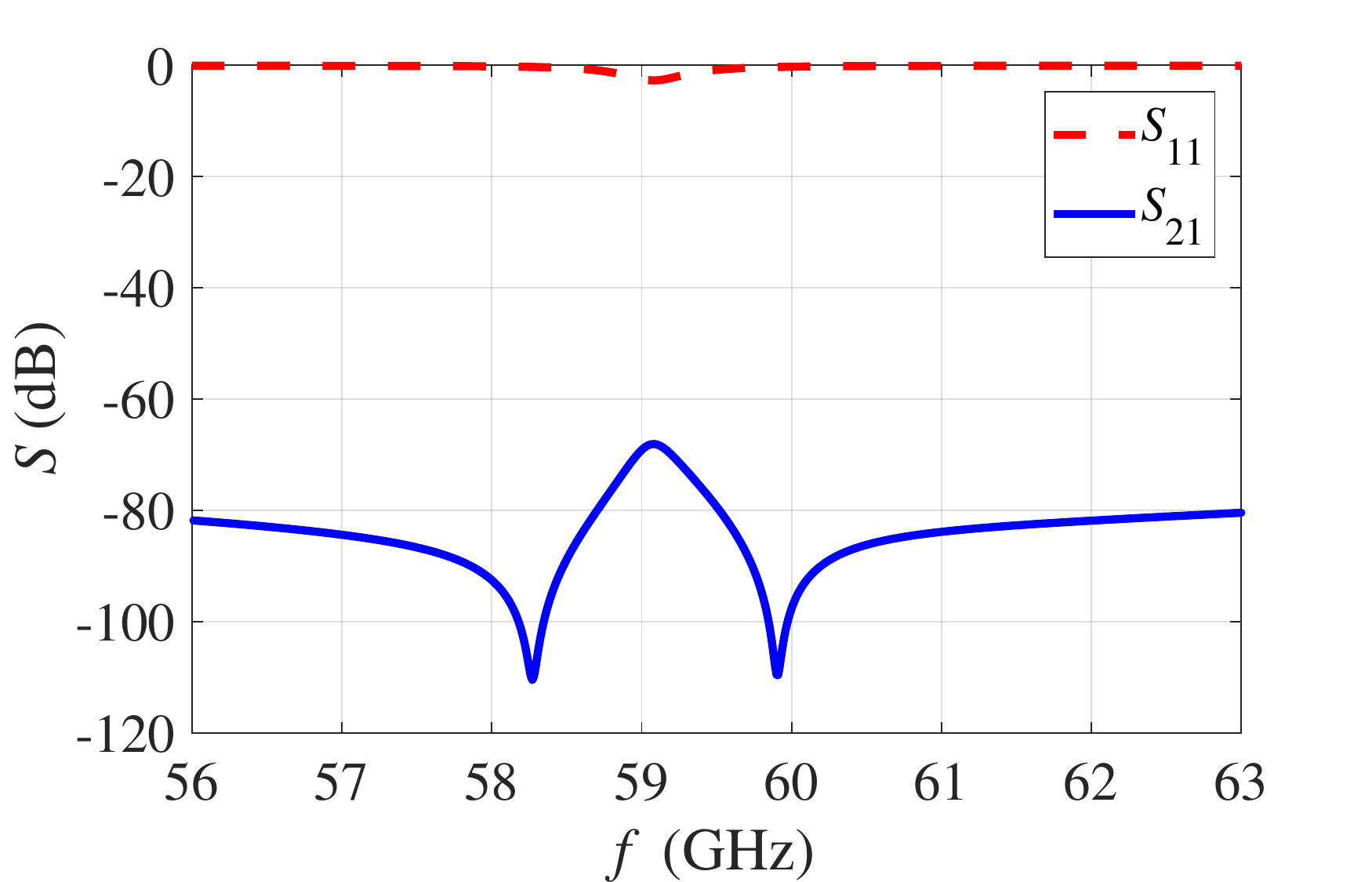}
	\caption{Scattering parameters of the structure shown in Fig.~\ref{fig:Fig5} in the ``off'' state. Geometrical parameters: $l=3.4$~mm, $p_{\rm{guide}}=1.5$~mm, $h_{\rm{guide}}=0.2$~mm $d_{\rm{guide}}=0.4$~mm, $p_{\rm{EBG}}=3$~mm, $h_{\rm{EBG}}=1.2$~mm $d_{\rm{EBG}}=0.8$~mm, $g=0.005$~mm. }
	\label{fig:S_off}
\end{figure}

Fig.~\ref{fig:S_off} displays the $S$ parameter results of the structure in Fig.~\ref{fig:Fig5} when the gap $g=g_\text{off}=0.005$~mm is considered (the ``off'' state discussed in the previous section). The result shows an insertion loss higher than 68 dB. Therefore, this waveguide can implement a switch with satisfactorily high isolation levels. %, a very low $S_{21}$ is needed in the matching bandwidth when the structure is in the ``off'' state.

\begin{figure}[!tbp]
    \includegraphics[width=\columnwidth]{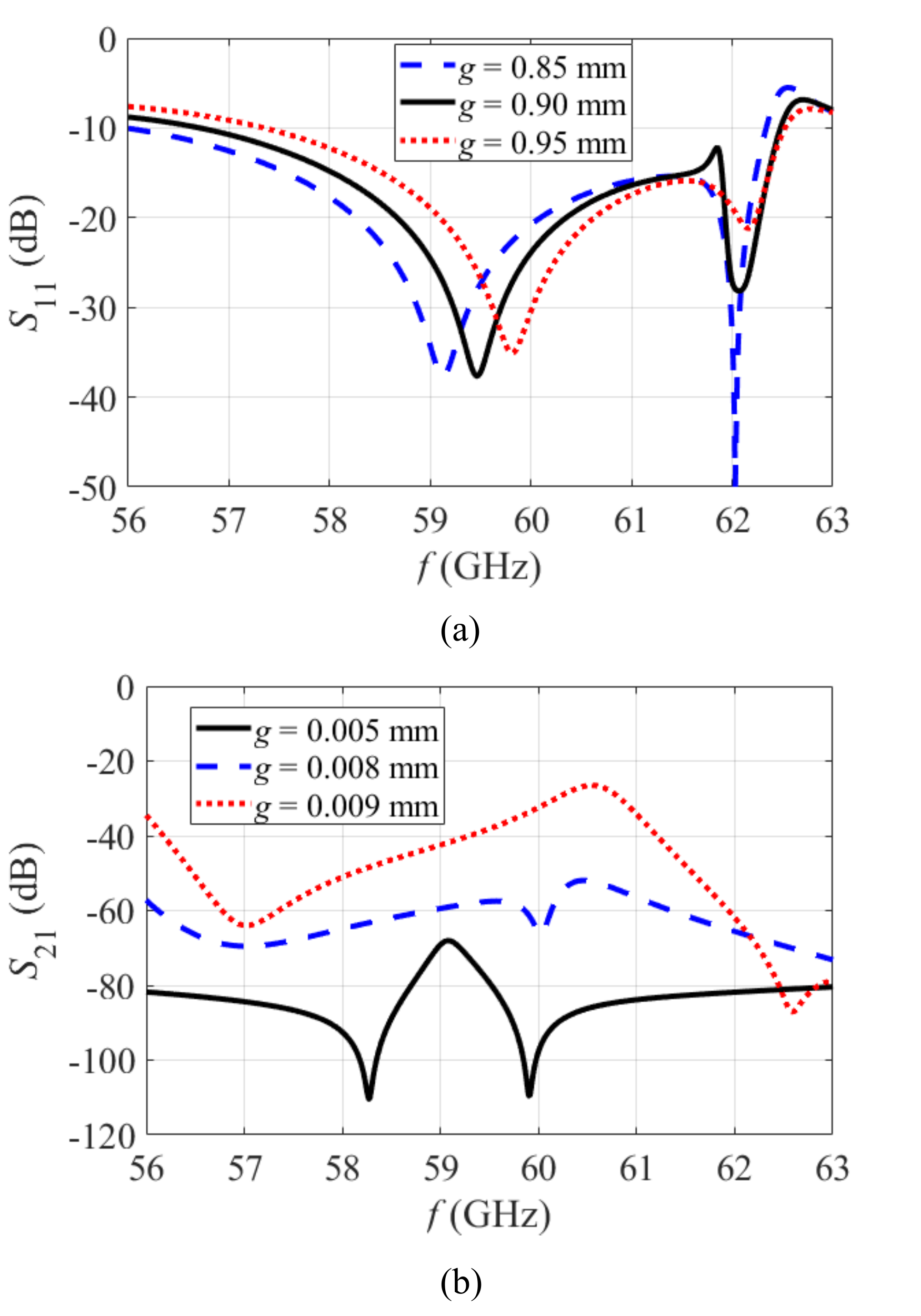}
	\caption{ (a) Scattering parameters of the structure shown in Fig.~\ref{fig:Fig5} for 3 different values of $g$ in the ``on'' state. (b) Scattering parameters of the structure shown in Fig.~\ref{fig:Fig5} for 3 different values of $g$ in the ``off'' state. Geometrical parameters: $l=3.4$~mm, $p_{\rm{guide}}=1.5$~mm, $h_{\rm{guide}}=0.2$~mm $d_{\rm{guide}}=0.4$~mm,  $p_{\rm{EBG}}=3$~mm, $h_{\rm{EBG}}=1.2$~mm $d_{\rm{EBG}}=0.8$~mm. }
	\label{fig:Fig_g}
\end{figure}

An important factor in the designed technology is the precision of the mechanical adjustments needed to achieve reconfigurability. This precision can be known by studying the sensitivity of the design to changes in the gap $g$. Fig.~\ref{fig:Fig_g}~(a) depicts $S_{11}$ results for the structure shown in Fig.~\ref{fig:Fig5} with $g=0.9$~mm in the ``on'' state and compares it to the two cases for which there is a $\Delta g=\pm0.05$~mm difference. The results show that such a change does not deteriorate the bandwidth by much and it only moves the $-10$~dB bandwidth to slightly higher or lower frequencies in each case. Fig.~\ref{fig:Fig_g}~(b) plots $S_{21}$ results for the same structure with $g=0.005$~mm in the ``off'' state and compares it to $g=0.008$~mm and $g=0.009$~mm. Here, we observe that the gap change of $\Delta g=+0.003$~mm does not jeopardize the design since the $S_{21}$ levels are kept below $-40$~dB after this change. The gap change of $\Delta g=+0.004$~mm provides $S_{21}$ levels that are lower than $-20$~dB.

\subsection{Adjacent Waveguides}
\label{subsec:2guides}
\begin{figure}[!tbp]
    \includegraphics[width=\columnwidth]{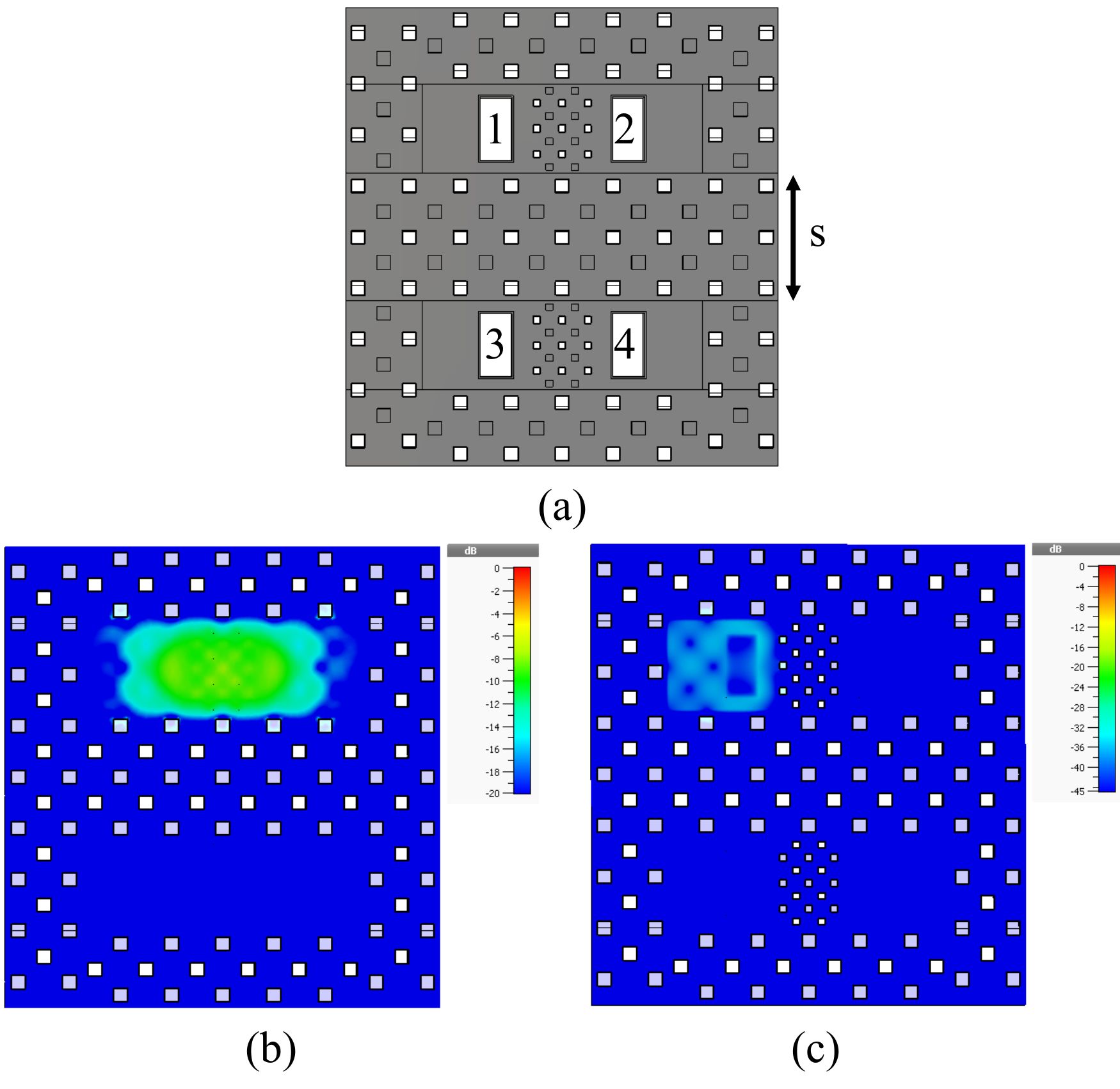}
	\caption{Two coupled lines: (a) the top view with the top metal plate removed for better visualisation of the inner parts (b) power flow while exciting port 1 at 60 GHz in "on" state (c) power flow while exciting port 1 at 60 GHz in "off" state.}
	\label{fig:Fig13}
\end{figure}

In this subsection, we study two adjacent waveguides to evaluate the coupling between the two waveguides, both in the ``on'' and ``off'' states, to provide some insights regarding the use of such technology in a switching network. Fig~\ref{fig:Fig13}~(a) displays the top view of the two adjacent waveguides with the top plate removed to better visualize their inner structure. The waveguide ports are numbered in the figure. We compute the power flow inside the two waveguides by exciting waveguide port 1. The waveguides are considered to be made of copper in the simulations. The distance between the two waveguides is 7.5~mm ($1.5\lambda_0$) and the length of the simulated waveguides is 25.5~mm (approximately $5\lambda_0$) where $\lambda_0$ is the wavelength of the electromagnetic wave in free space at 60~GHz. Fig~\ref{fig:Fig13}~(b) displays the power flow inside the waveguides at 60 GHz when the structure is in the ``on'' state. It is clearly seen from the picture that there is a power flow along the longitudinal direction of the excited waveguide from port 1 towards port 2. It is also clearly visible that there is a good isolation between the two waveguides and that the EBG is confining the fields inside the upper waveguide. In contrast, Fig~\ref{fig:Fig13}~(c) shows the power flow at 60
~GHz when the structure is in the ``off'' state. We observe that there is no power flow along the length of the top excited waveguide. The waveguide is attenuating the wave in both cross-sectional and longitudinal directions, therefore preventing propagation of the wave in any direction.

\begin{figure}[!tbp]
    \includegraphics[width=\columnwidth]{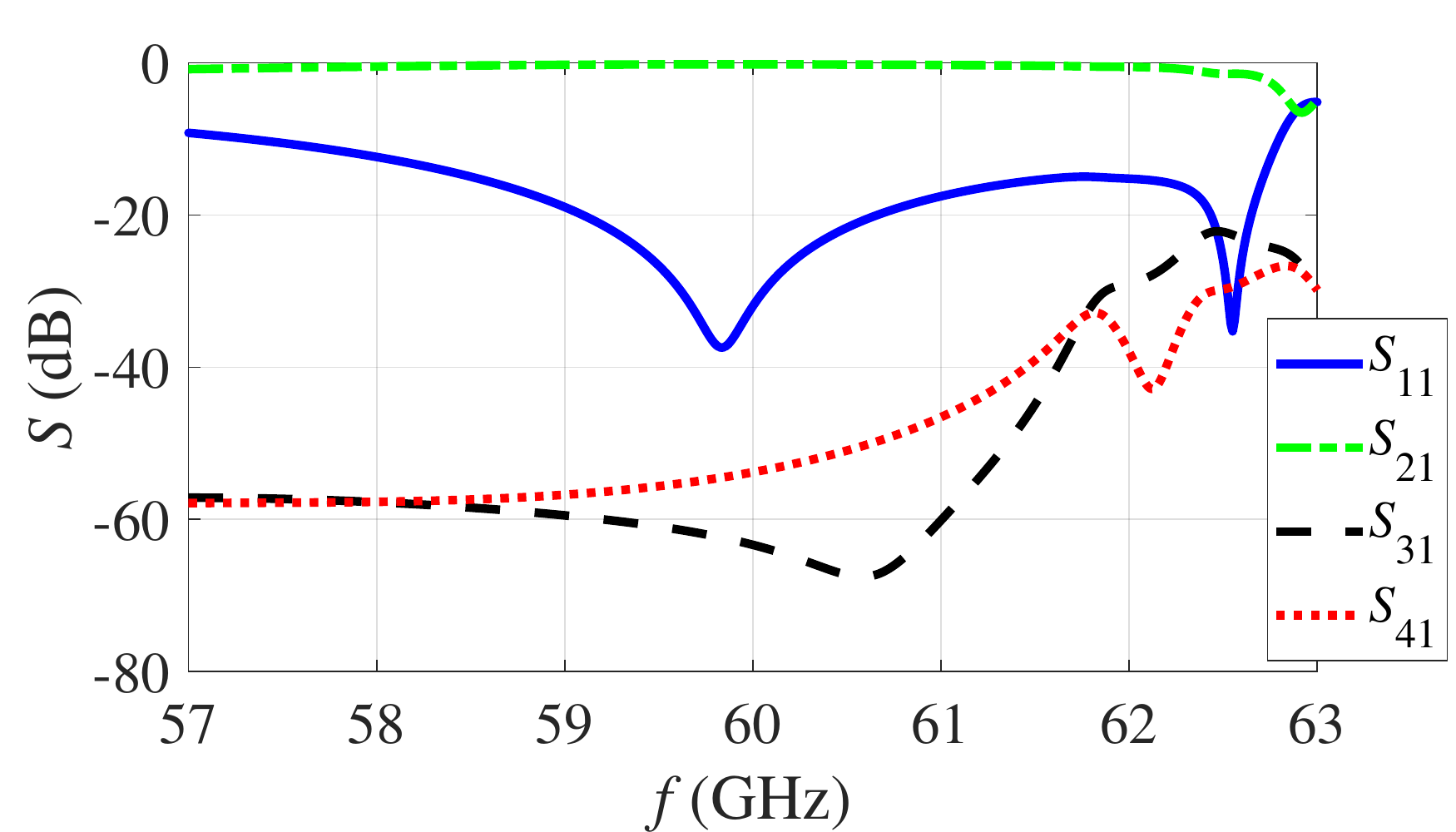}
	\caption{Scattering parameters of the structure shown in Fig.~\ref{fig:Fig13}~(a).}
	\label{fig:Fig14}
\end{figure}

Fig.~\ref{fig:Fig14} plots the scattering parameters of the structure in Fig.~\ref{fig:Fig13}~(b) by exciting the upper waveguide through waveguide port 1. $S_{11}$ and $S_{21}$ curves are similar to those shown in Fig.~\ref{fig:Fig7} and show that the waveguide is still functioning correctly in its bandwidth. The -10~dB $S_{11}$ bandwidth is 57.4-62.8~GHz. The insertion loss of less than 0.7 dB is achieved for most of the $S_{11}$ bandwidth (57.4-62.2 GHz). In a small portion of the $S_{11}$ bandwidth (62.2-62.8 GHz) the insertion loss goes beyond 0.7 dB and increases to a maximum of 3.3 dB at the end of the $S_{11}<-10$~dB bandwidth. In addition to these two curves, $S_{31}$ and $S_{41}$ results are also plotted to provide the coupling levels between ports of different waveguides. We observe that there is a good isolation between the waveguides and that the coupling is below -22~dB over the whole operational bandwidth.

% for old picture: coupling is studied in more detail in Fig.~\ref{fig:Fig15}~(a), which displays $S_{31}$ for different values of lateral spacing $s$ between the two adjacent waveguides shown in Fig.~\ref{fig:Fig13}~(a). It is observed that the $S_{31}$ decreases as the distance between the two waveguides is increased. With an increased distance, the wave is more attenuated across the EBG region between the two waveguides. An $s=1.5p_{\textrm{EBG}}$ results in $S_{31}$ of about $-40$~dB in the center of the $S_{11}$ bandwidth previously discussed, and up tp -20~dB on the edge of this bandwidth. Distances of $s=2.5p_{\textrm{EBG}}$ and $s=3.5p_{\textrm{EBG}}$ provide better isolation as their $S_{31}$ is below -40~dB for most of the $S_{11}$ bandwidth. The case $s=4.5p_{\textrm{EBG}}$ has $S_{31}$ levels below -40~dB in the entire $S_{11}$ bandwidth. Here $p_{\textrm{EBG}}=3$~mm, so $1.5p_{\textrm{EBG}}$ is slightly smaller than the free space wavelength $\lambda_0=5$~mm at 60 GHz. Fig~\ref{fig:Fig15}~(b) plots $S_{41}$ for different values of spacing $s$ to study the effect of $s$ on this result. The findings are similar to those from $S_{31}$. An increase of $s$ leads to a decrease of $S_{41}$. Again, to achieve $S_{41}$ lower than -40 dB in all the -10 dB $S_{11}$ bandwidth, a minimum distance of $s=4.5p_{\textrm{EBG}}$ is required.

\begin{figure}[!tbp]
    \includegraphics[width=\columnwidth]{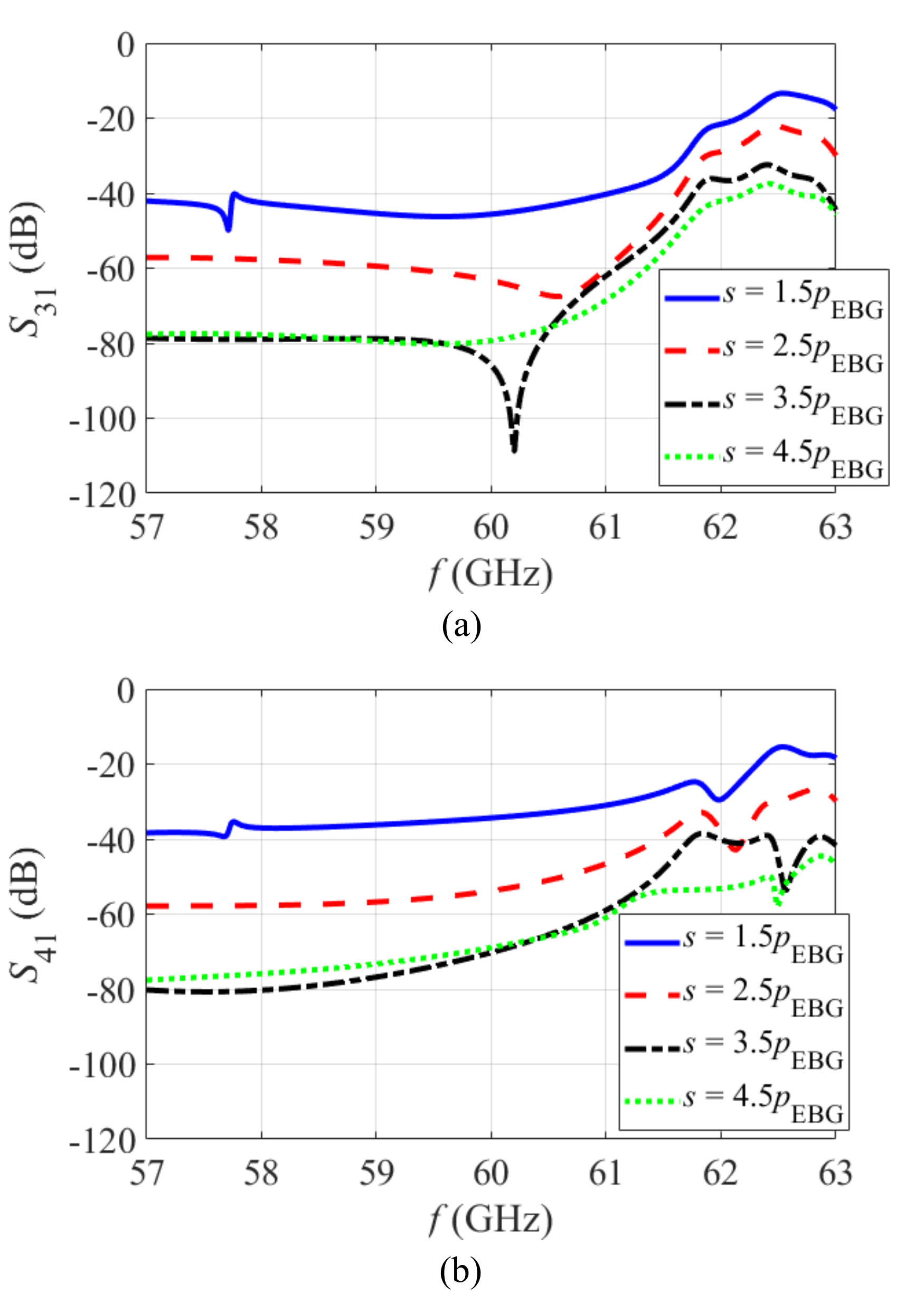}
	\caption{(a) $S_{31}$ in the structure shown in Fig.~\ref{fig:Fig13}~(a) for different values of $s$ (b) $S_{41}$ in the structure shown in Fig.~\ref{fig:Fig13}~(a) for different values of $s$.}
	\label{fig:Fig15}
\end{figure}

The coupling is studied in more detail in Fig.~\ref{fig:Fig15}~(a), which displays $S_{31}$ for different values of lateral spacing $s$ between the two adjacent waveguides shown in Fig.~\ref{fig:Fig13}~(a). It is observed that the $S_{31}$ decreases as the distance between the two waveguides is increased. With an increased distance, the wave is more attenuated across the EBG region between the two waveguides.  An $s=1.5\:p_{\textrm{EBG}}$ results in $S_{31}$ of below $-40$~dB in the first 4 GHz of the $S_{11}$ bandwidth previously discussed. But it increases up to -13.5~dB at one point in the $S_{11}$ bandwidth. Distances of $s=2.5\:p_{\textrm{EBG}}$ and $s=3.5\:p_{\textrm{EBG}}$ provide better isolation as their $S_{31}$ is below -40~dB for a larger part of the $S_{11}$ bandwidth. $S_{31}$ levels for $s=2.5\:p_{\textrm{EBG}}$ and $s=3.5\:p_{\textrm{EBG}}$ are respectively below -22 dB and -32 dB over the whole bandwidth. The case $s=4.5\:p_{\textrm{EBG}}$ has $S_{31}$ levels below -37~dB in the entire $S_{11}$ bandwidth. Here $p_{\textrm{EBG}}=3$~mm, so $1.5\:p_{\textrm{EBG}}$ is ten percent smaller than the free space wavelength $\lambda_0=5$~mm at 60 GHz. Fig~\ref{fig:Fig15}~(b) plots $S_{41}$ for the same four values of spacing $s$ to study the effect of $s$ on this result. The findings are similar to those from $S_{31}$. An increase of $s$ leads to a decrease of $S_{41}$. Again, $s=4.5\:p_{\textrm{EBG}}$ achieves the lowest $S_{41}$ levels which is lower than -44 dB in all the -10 dB $S_{11}$ bandwidth.

\section{Conclusion}
\label{sec:conclusion}
An artificial waveguide technology using glide symmetry is introduced. The use of this symmetry in the guiding medium in addition to the EBG material provides a dispersionless propagation of the electromagnetic wave. Furthermore, the periodicity in the guiding region allows for the transition from the passband to a stopband in the 57.4-62.2~GHz frequency band. This transition can be achieved by a mechanical reconfiguration of the structure, namely readjusting the displacement between the two metallic plates. This mechanical reconfiguration of the glide-symmetric waveguide has been shown to be an effective way to achieve a two-state on-off structure exhibiting in the off-state, an isolation better than 65~dB and in the on-state, insertion losses less than 0.7~dB. This technology may find applications in the design of low-loss RF switches.

As an important feature of the proposed technology, the signal propagates through air and not through any semi-conductor or dielectric material. So, the only losses are the ohmic losses of the metallic structure and the proposed approach is therefore able to handle high power. This makes glide-symmetric waveguides proper candidates to switch between multiple input ports of MMW multibeam antennas for instance.

% or


%\appendices

%\section{}

%\section*{Acknowledgment}


\ifCLASSOPTIONcaptionsoff
  \newpage
\fi

\bibliographystyle{IEEEtran}
\bibliography{IEEEabrv,References}

\end{document}